\documentclass[prb,twocolumn]{revtex4}
\usepackage[dvips]{color}
\usepackage[dvips]{psfrag}
\usepackage{lscape,graphicx}
\usepackage{subfig}
\usepackage{CJK}
\usepackage{caption}
\usepackage{ellipsis}
\usepackage{boxedminipage}
\usepackage{amsbsy,amssymb,amsmath,bm}
\usepackage{url}
\begin{document}
\newcommand{\LMO}{LaMnO$_3$}
\newcommand{\SMO}{SrMnO$_3$}
\newcommand{\NMO}{NdMnO$_3$}
\newcommand{\LAO}{LaAlO$_3$}
\newcommand{\STO}{SrTiO$_3$}

\title{Relaxor characteristics at the interfaces of [NdMnO$_3$/SrMnO$_3$/LaMnO$_3$] superlattices}
\begin{CJK*}{KS}{}
\author{Jiwon Seo$^{1,6}$, Bach T. Phan$^{2,3}$, Jochen Stahn$^4$, Jaichan Lee$^2$, and Christos Panagopoulos$^{1,5,6}$}
\affiliation{$^1$Cavendish Laboratory, University of Cambridge, Cambridge CB3 0HE, UK}
\affiliation{$^2$School of Advanced Materials Science and Engineering, Sungkyunkwan University, Suwon, South Korea}
\affiliation{$^3$Faculty of Materials Science, University of Science, Vietnam National University, Vietnam}
\affiliation{$^4$ETH Zurich \& Paul Scherrer Institut Laboratory for Neutron Scattering, Switzerland}
\affiliation{$^5$Department of Physics, University of Crete and FORTH, 71003 Heraklion, Greece}
\affiliation{$^6$Division of Physics and Applied Physics, Nanyang Technological University, 637371 Singapore}

\begin{abstract}
We have investigated the magnetic properties of transition metal oxide superlattices with broken inversion symmetry composed of three different antiferromagnetic insulators, [NdMnO$_3$/SrMnO$_3$/LaMnO$_3$]. In the superlattices studied here, we identify the emergence of a relaxor, glassy-like behavior below T$_{SG}$ =36K. Our results offer the possibility to study and utilize magnetically metastable devices confined at nano-scale interfaces.
\end{abstract}
\maketitle
\end{CJK*}

Heterostructures of materials with strong electron-electron and electron-lattice interactions, the so-called correlated electron systems, are potential candidates for emergent interfacial properties including various forms of spin, charge and orbital ordering absent in bulk materials. Promising examples include multilayers composed of insulators of LaAlO$_3$ and SrTiO$_3$ with interfaces displaying properties of quasi-two-dimensional electron gases \cite{Thiel-2Dgas}, superconductivity \cite{Reyren_superconducting}, metallic conductivity \cite{LAOSTO_metal1, LAOSTO_metal2, LAOSTO_metal3} and ferromagnetism (FM) \cite{STOLAO_magnetic}. The multilayers composed of antiferromagnetic (AF) insulators in bulk forms, \LMO\ and \SMO\ are also examples of emergent electromagnetic properties at the interfaces between dissimilar manganites \cite{LaMnOSrMnO_PNR, LaMnOSrMnO_2002, magnetism conductivity and orbital order in (LaMnO3)2n/(SrMnO3)n superlattices, Viscous spin exchange torque on preessional magnetization in (LaMnO3)2n/(SrMnO3)n superlattices, Phse evolution and critical behavior in strain-tuned LaMnO-SrMnO superlattices,Electronic Reconstruction at SrMnO3-LaMnO3 Superlattice Interfaces}. It has been demonstrated that these superlattices could posses FM order at the interfaces due to a charge reconstruction, although each parent material is an AF \cite{Electronic Reconstruction at SrMnO3-LaMnO3 Superlattice Interfaces}. The competitive interaction between the reconstructed FM at interfaces and the AF states present far from the interface regions has been suggested to lead to a frustrated/glass like behavior \cite{Viscous spin exchange torque on preessional magnetization in (LaMnO3)2n/(SrMnO3)n superlattices, metal insulator transition and its relation to magnetic structure}.
Small external perturbation in glass-like correlated electron thin film devices, at a ``caged" nano-structured interface, in  particular, is expected to lead to high degree of tunability. These include, magnetoelectronics such as spin and charge memory devices at the atomic scale.




\begin{figure}
\includegraphics[scale=0.6]{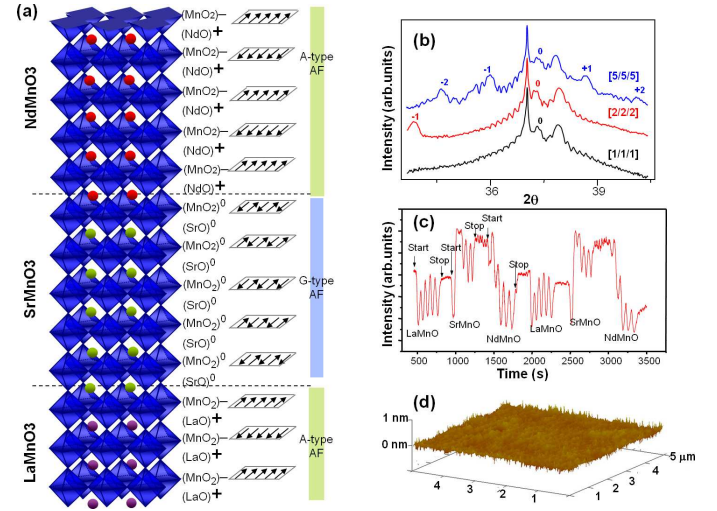}
\caption{\label{fig1} (a) Schematic drawing of the superlattice
[(NdMnO$_{3}$)$_5$/(SrMnO$_{3}$)$_5$/(LaMnO$_{3}$)$_5$]$_{8}$. The octahedral structures and spheres represent BO$_6$
and A-site atoms in the ABO$_3$ perovskite structure, respectively. The arrays of the arrows represent corresponding antiferromagnetic types. (b) Synchrotron x-ray diffraction for different superlattices. (c) In-situ
reflection high energy electron diffraction for the superlattice with $n$=5. (d) The topography image of
atomic force microscopy for the superlattice with $n$=5. }
\end{figure}

Here we report on the relaxor and spin glass-like properties arising at the interface of superlattices, composed of insulating manganites : \LMO, \SMO\ and \NMO, which are A-, G, and A-type AF, respectively.
Superlatties of [(NdMnO$_{3}$)$_n$/(SrMnO$_{3}$)$_n$/(LaMnO$_{3}$)$_n$]$_m$ were
grown epitaxially on single crystalline SrTiO$_{3}$ substrates at
an ambient oxygen/ozone mixture of 10$^{-4}$ Torr by layer-by-layer growth technology using the laser
molecular beam epitaxy technique. The details were reported in an earlier work \cite{tricolor_growing}. Figure 1 (a) depicts a schematic drawing of a superlattice with $n$=5 studied here, with alternate A
sites around the blue octahedra representing MnO$_6$ in the
ABO$_3$ pervoskite. The total thickness of the superlattices was
kept approximately 500 \AA {}, varying ($n$, $m$) = (1 unit cell, 42), (2,
21), (5, 8), and (12, 4) in order to investigate the effect of the period.
Structural characterization
using synchrotron x-ray diffraction (Fig. 1(b)) along with the in-situ
reflection high energy electron diffraction (Fig. 1(c)) indicate the
presence of sharp interfaces with roughness
less than one unit cell. The topography image performed using
atomic force microscopy (Fig. 1(d)) confirms the surface
roughness to be less than one unit cell.
\begin{figure}
\includegraphics[scale=0.5]{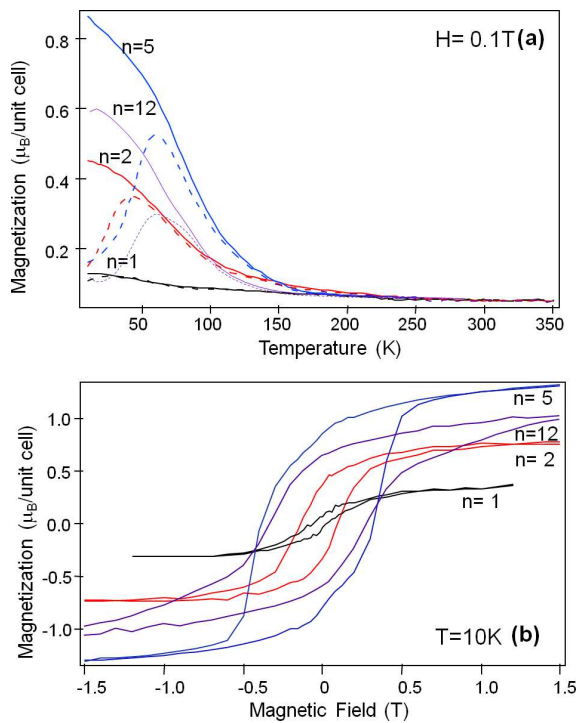}
\caption{\label{fig2} (a) Temperature dependence of the magnetization
measured in a magnetic field of 0.1T applied parallel to the
plane of the films. (b) Hystereses
loops obtained at 10K with a magnetic field applied
parallel to the plane of the films.}
\end{figure}

The bulk magnetic properties of the superlattices were investigated using a superconducting quantum interference device magnetometer. Figure 2(a) shows magnetization curves as a function of temperature.  Data were taken by warming the sample in a field
(after cooling to 10K in zero-field) (ZFC: dashed lines), and by cooling in the presence of a field (FC: solid lines). The discrepancy between FC and ZFC curves at low temperature broadly resembles spin-glasses (SG). We will discuss this later. For superlattices with $n$$\geqq$2, the magnetization values are significantly larger (8 times larger in the cases of $n$=5) than those of bulk LaMnO$_{3}$ \cite{LaMnOSrMnO_2002}, SrMnO$_{3}$ \cite{LaMnOSrMnO_2002} and NdMnO$_{3}$
\cite{NdMnO3}. On the other hand, for $n$=1 there is a weak magnetic moment which is comparable to that of bulk LaMnO$_{3}$, SrMnO$_{3}$, or NdMnO$_{3}$. The magnetic properties of (NdMnO$_{3}$)$_{1}$/(LaMnO$_{3}$)$_{1}$/(SrMnO$_{3}$)$_1$ may be similar to the solid solution states of (Nd, Sr, La)MnO$_3$ due to charge spreading through the interfaces, resulting in a three dimensional uniform charge distribution while keeping chemically sharp interfaces \cite{magnetism conductivity and orbital order in (LaMnO3)2n/(SrMnO3)n superlattices}. The weaker magnetization observed for the superlattice with $n$=1 compared to (LaMnO$_3$)$_2$/(SrMnO$_3$)$_1$, is due to a decreased magnetization caused in a La$_{0.7}$Sr$_{0.3}$MnO$_3$ solid solution by replacing the La by  Nd, Pr or Y which have smaller ionic radii  \cite{Lattice effects on the magnetoresistance in doped LaMnO3}. The tendency for an increase in magnetization and Curie Temperature with increasing period until a critical period, $n$=5, and a decrease with an increase of n above 5 (Fig. 2(a)) agrees with earlier suggestions for [(LaMnO$_{3}$)$_{n}$/(SrMnO$_{3}$)$_n$] \cite{LaMnOSrMnO_2002}.

 Hystereses loops (Fig. 2 (b)) were measured at 10K after field cooling in 0.1T applied along the plane of the film. (The linear part of the hysteresis due to the paramagnetic substrate has been subtracted.) The coercive fields of the samples are $H_c$=0.04, 0.14, 0.36 and 0.28T for $n$=1, 2, 5 and 12, respectively. The coercive fields along with magnetic moments reveal a critical period of $n$=5, indicating the presence of FM phases as previously reported for a similar superlattice of [LaMnO$_{3}/$SrMnO$_{3}$] \cite{LaMnOSrMnO_2002, magnetism conductivity and orbital order in (LaMnO3)2n/(SrMnO3)n superlattices,LaMnOSrMnO_PNR,Viscous spin exchange torque on preessional magnetization in (LaMnO3)2n/(SrMnO3)n superlattices}.


To further investigate the presence of the thermal hysteresis at low temperatures (Fig. 2(a)), we examined
the sample with $n$=2 by measuring the dc magnetic susceptibility ($\chi$) as a function of temperature in
different magnetic fields (0.05T -- 1.5T).  In Fig. 3 the dashed and solid lines depict the susceptibility obtained in ZFC and FC, respectively. The shift of the peaks of the ZFC curves to lower temperatures with increasing field is a characteristic of SG/relaxors. The normalized spin glass order parameter $q$ is defined as \cite{scaling}
\begin{equation}\label{q}
    q(T,H)=[(\chi_0+C/T)-
\chi(T,H)]/(C/T)
\end{equation}
or
\begin{equation}\label{scaling_equation}
q(T,H) =|t|^\beta F_{\pm}(H^2/|t|^{\beta+\gamma})
\end{equation}
where C, F$_{\pm}$, $t$, $\beta$, and $\gamma$ are the Curie constant,
the scaling function, the reduced temperature
$t$=(T-T$_{SG}$)/T$_{SG}$  (here T$_{SG}$ is the SG temperature), and the
critical exponents characterizing the SG behavior, respectively.
Through the scaling analysis (Fig. 3 (inset))
we obtain $T_{SG}$=36K, $\beta$=0.7 and
$\gamma$=1.95. These values are
in good agreement with experimental reports for other SG such as CdIn$_{0.3}$Cr$_{1.7}$S$_4$
($\beta$=0.75 and $\gamma$=2.3 \cite{CdInCrS}).
Notably there is deviation from the scaling function at low fields (0.05 and 0.1T). This behavior may be due to an inhomogeneous SG order in the superlattices, such as coexistance of the former with AF and FM regions whose volume ratio may be changed by an applied magnetic field.
For samples with $n\geqq$5 we do not observe the scaling law because the coexistence and modulation of the SG, AF, and FM phases as a function of thickness hinders the characterization of the SG behavior from the other regions. 

\begin{figure}
\includegraphics[scale=0.6]{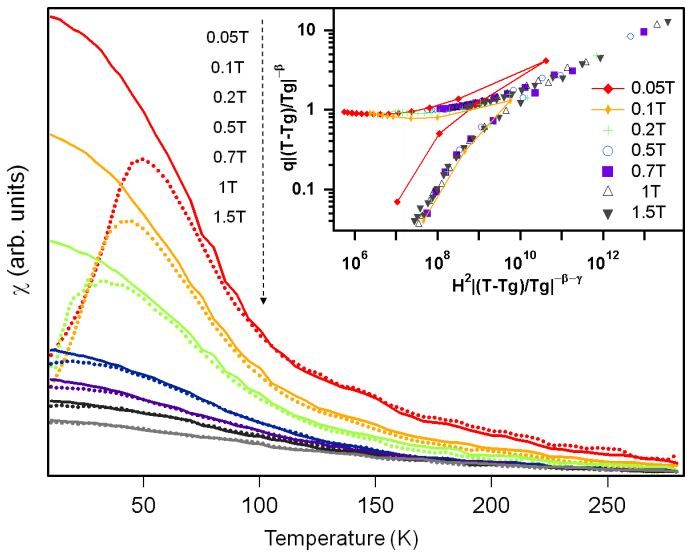}
\caption{\label{fig3} Magnetic susceptibility ($\chi$) for magnetic fields from 0.05T to 1.5T for the sample with $n$=2. The dashed and solid lines depict the susceptibility obtained with ZFC and FC conditions, respectively.  (Inset) Spin glass scaling for the superlattice with $n$=2.}
\end{figure}

\begin{figure}
\includegraphics[scale=0.57]{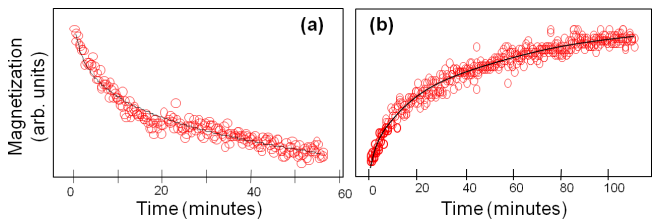}
\caption{\label{aging} Time response of the magnetization for
[(\NMO)$_2$/(\SMO)$_2$/(\LMO)$_2$]. (a) The relaxation of the
magnetization is measured at 10K after cooling in a magnetic field of 0.1T applied parallel to the plane of the film. The solid line is a fit to  $M(t)$ = $M_0$exp(-$\alpha(t/\tau_0)^{1-y}/(1-y)$) (b) The increase of the magnetization is measured at 10K in a field of 0.1T applied parallel to the film's plane after cooling in zero field. The solid line is a fit to $M(t)$ = $M_0$ + $Sln(t+t_0)$.}
\end{figure}

The time decay of the magnetization for the superlattice with $n$=2 (Fig. 4) adds credence to the glassy characteristics. (The other films also show time relaxation but we do not present the data here.) The relaxation of the thermoremanent magnetization (Fig. \ref{aging} (a)) was measured by the following method. The sample was cooled from room temperature to 10K in the presence of a magnetic field of 0.1T applied parallel to the film's plane. When the temperature was stable, the magnetic field was switched off and the magnetization decay was recorded as a function of time for 60 min. The decay curve is fitted by a stretched-exponential function (solid line) \cite{1-n} $M(t) = M_0exp(-\alpha(t/\tau_0)^{1-y}/(1-y))$.
We find $y$ =0.7 which is typical of other SG systems such as AgMn \cite{1-n}.
The slow increase of the magnetization after switching on the magnetic field is depicted in Fig. 4(b). Here the sample with $n$=2 was cooled down from room temperature to 10K in the absence of a magnetic field. When at 10K, a field of 0.1T was applied parallel to the surface of the film and data was recorded. The data is fitted by the logarithmic function (solid line) \cite{SG_aging} $M(t) = M_0 + Sln(t+t_0)$.

The thermal hysteresis (Fig. 2(a)), the scaling curve (Fig. 3: inset) and the ageing signatures (Fig. 4) reveal the presence of a SG like behavior. Possible origins for SG in this system include : (1) SG characteristics present in each layer, (2) miscut between substrate and the first layer from the substrate, LaMnO$_3$, and (3) magnetic frustration between FM and AF regions, where FM and AF regions are possibly present at interfaces and in the core of each layer, respectively. The possibility for this behavior being due to each layer can be ruled out, however, by the systematic changes in the amplitudes and the irreversible temperatures of the magnetization for samples with different periods (Fig. 2). Also magnetization curves of each individual manganite show AF not a SG. A possible source for the SG characteristics may be the interface between the substrate (SrTiO$_3$) and the first deposited layer of LaMnO$_3$. This too, cannot be the origin for our observations, since the magnetization curves as a function of temperature for a 60 unit cells LaMnO$_3$ layer grown on SrTiO$_3$ indicates AF not a SG \cite{Enhanced ordering temperatures in antiferromagnetic manganite superlattices}. 
We believe a competition between the FM  and AF layers may account for our observations. In fact, such a competition has already been proposed for superlattices of [(LaMnO$_{3}$)$_{2n}$/(SrMnO$_{3}$)$_n$] \cite{Viscous spin exchange torque on preessional magnetization in (LaMnO3)2n/(SrMnO3)n superlattices, Phse evolution and critical behavior in strain-tuned LaMnO-SrMnO superlattices}.

The modulated magnetization of AF and FM layers as a function of depth was studied using polarized neutron reflectivity (PNR) in a superlattice with $n$=12, whose period is most suitable for PNR. Figure 5(a) shows the PNR results at 300K (above T$_c$) with non-polarized neutrons since there is no magnetic signature at this temperature. The solid line depicts a fit of the calculated reflectivity obtained from the scattering length density (SLD) model as a function of depth. The SLD profile (inset) for NdMnO$_3$, \SMO, and \LMO\ gives 3.65, 3.55 and 3.75$\times$10$^{-6}$ \AA$^{-2}$, respectively, in good agreement with calculated and experimental values \cite{LaMnOSrMnO_PNR}. The weak Bragg peak at $q$ = 0.045 \AA$^{-1}$ is due to the similarity in the nuclear scattering length for La, Sr and Nd atoms. The reflectivity measured in a magnetic field of 0.6T applied parallel to the film's surface, after field cooling to 10K in 0.6T, shows strong Bragg peaks and significant difference between R$^+$ and R$^{-}$, indicating the presence of a magnetic modulation in the superlattice. R$^+$ and R$^{-}$ are obtained by the polarized neutrons with spin states parallel and antiparallel to the magnetic field, respectively. From our best fit to the PNR data, we obtained the magnetic profile shown in the inset of Fig. 5. As in earlier reports on superlattices composed of \SMO\ and \LMO\ layers \cite{LaMnOSrMnO_PNR}, our data also reveal an enhancement in the magnetization at the interfaces of \NMO/\SMO\ (1.1$\mu_B$/unit cell) and \SMO/\LMO\ (3.3$\mu_B$/unit cell). The obtained thickness of the interfaces is around 10 \AA. Notably, there is no signature of an enhancement at interfaces of \LMO/\NMO. This may be due to the absence of polarity-discontinuity between these layers \cite{LAOSTO_metal1}. In the regions far from the interfaces, \NMO\ (0.7$\mu_B$/unit cell), \SMO\ ($<$0.1$\mu_B$/unit cell), and \LMO\ (1.5$\mu_B$/unit cell) layers have comparable values to single films grown on \STO\, or in bulk \cite{Enhanced ordering temperatures in antiferromagnetic manganite superlattices, NdMnO3, LaMnOSrMnO_PNR}. An integrated magnetization estimated from the values we obtained by the fitting in Fig. 5 for the film with $n$=12 is within 10\% of the saturated magnetization moment obtained by bulk magnetization (Fig. 2(b)). We assumed that the magnetization and the thickness of the interfaces for the film with $n$=5 are same to those values obtained from the films with $n$=12. An integrated magnetization for the film with $n$=5 which is obtained based on the above-mentioned assumptions is also within 10\% of the value obtained by bulk magnetization. Therefore, we may interpret the relaxor behavior being due to the competitive interaction between FM mainly present at interfaces and AF regions in a magnetically modulated system \cite{interface_SG_Fe_FeOxide, LFOLMO}. The large coercive fields (Fig. 2(b)) commonly occurring by pinning the FM spins nearby an AF layer support the competition at FM/AF interfaces. 

\begin{figure}
\includegraphics[scale=0.58]{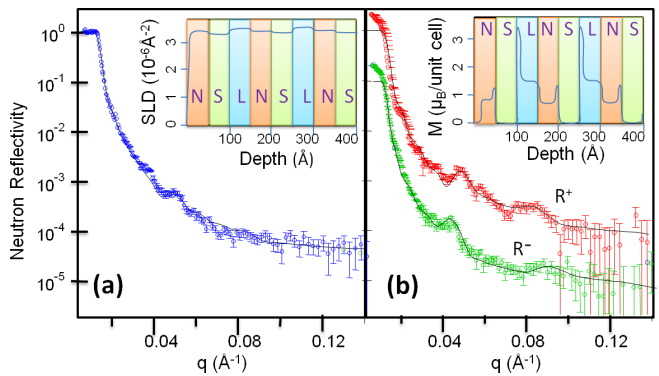}
\caption{\label{PNR} (Color online) (a) Polarized neutron reflectivity (PNR) measurements for the sample with $n$=12 taken at 300K (above T$_C$). The orange, green and blue regions depict the regions of \NMO(N), \SMO(S), and \LMO(L) layers, respectively. The solid line is a fit of the calculated reflectivity obtained using the scattering length density (SLD) model. (b) PNR taken at 10K ( below T$_C$) in 0.6T after field cooling in the same field. The red (upper) and green (lower) circles are the R$^+$ and R$^-$ data, respectively. (Inset) the magnetic structure obtained from a calculation which reproduce the PNR data as represented with the solid lines.}
\end{figure}

In summary, we fabricated a series of superlattices stacked repeatedly by different types of AF insulators namely, LaMnO$_{3}$, SrMnO$_{3}$ and NdMnO$_{3}$. The magnetic properties obtained by bulk magnetometry have revealed the presence of FM, AF and spin glass (SG) phases. The thermal hysteresis and time dependent magnetization indicate a SG like behavior below T$_{SG}$ (=36K). Scaling shows the critical exponents to be $\beta$=0.7 and $\gamma$=1.95. The possible origin of the SG characteristics may be due to the competing interactions between FM and AF regions. A modulation of FM and AF regions have been detected by polarized neutron reflectivity. This study may be potentially applicable to metastable magnetic memory devices which can offer a gateway to engineer sub-nano-scale metastates confined at oxide interfaces.

\section*{ACKNOWLEDGEMENTS}
This work is supported by The Royal Society,  EURYI, MEXT-CT-2006-039047, Korea Research Foundation Grant (KRF-2005-215-C00040), the Basic Research Program (2009-0092809) through the National Research Foundation of Korea, and the National Research Foundation of Singapore.


\begin{thebibliography}{99}

\bibitem{Thiel-2Dgas} S. Thiel, G. Hammerl, A. Schmehl, C. W. Schneider and J. Mannhart, \textit{Science} \textbf{313}, 1942 (2006)


\bibitem{Reyren_superconducting} N. Reyren, S. Thiel, A. D. Caviglia, L. Fitting Kourkoutis, G. Hammerl, C. Richter, C. W. Schneider, T. Kopp, A.-S. Ruetschi, D. Jaccard, M. Gabay, D. A. Muller, J.-M. Triscone and J. Mannhart, \textit{Science} \textbf{317}, 1196 (2007)

\bibitem{LAOSTO_metal3}W. Siemons, G. Koster, H. Yamamoto, T. H. Geballe, D. H. A. Blank and M. R. Beasley, Phys. Rev. B 76, 155111 (2007)

\bibitem{LAOSTO_metal1} A. Ohtomo and H. Y. Hwang, \textit{Nature}, \textbf{427}, 423 (2004)

\bibitem{LAOSTO_metal2} M. Huijben, G. Rijnders, D. H. A. Blank, S. Bals, S. Van Aert, J. Verbeeck, G. Van Tendeloo, A. Brinkman and H. Hilgenkamp , \textit{Nature Materials}, \textbf{5}, 556 (2006)

\bibitem{STOLAO_magnetic} A. Brinkman, M. Huijben, M. van Zalk, J. Huijben, U. Zeitler, J. C. Maan, W. G. van der Wiel, G. Rijnders, D. H. A. Blank and H. Hilgenkamp, \textit{Nature materials}, \textbf{6}, 493 (2007)


\bibitem{Phse evolution and critical behavior in strain-tuned LaMnO-SrMnO superlattices} H. Yamada, P. H. Xiang and A. Sawa, \textit{Phys. Rev. B}, \textbf{81}, 014410 (2010)
\bibitem{LaMnOSrMnO_2002} T. Koida, M. Lippmaa, T. Fukumura, K. Itaka, Y. Matsumoto, M. Kawasaki and H. Koinuma, \textit{Phys. Rev. B}, \textbf{66}, 144418 (2002)
\bibitem{magnetism conductivity and orbital order in (LaMnO3)2n/(SrMnO3)n superlattices} S. Dong, R. Yu, S. Yunoki, G. Alvarez, J.-M. Liu and E. Dagotto, \textit{Phys. Rev. B} \textbf{78}, 201102(R) (2008)
\bibitem{LaMnOSrMnO_PNR} S. J. May, A. B. Shah, S. G. E. te Velthuis, M. R. Fitzsimmons, J. M. Zuo, X. Zhai, J. N. Eckstein, S. D. Bader and A. Bhattacharya, \textit{Phys. Rev. B} \textbf{77} 174409 (2008)
\bibitem{Electronic Reconstruction at SrMnO3-LaMnO3 Superlattice Interfaces} S. Smadici, P. Abbamonte, A. Bhattacharya, X. Zhai, B. Jiang, A. Rusydi, J. N. Eckstein, S. D. Bader and Jian-Min Zuo, \textit{Phys. Rev. Lett.} \textbf{99} 196404 (2007)
\bibitem{Viscous spin exchange torque on preessional magnetization in (LaMnO3)2n/(SrMnO3)n superlattices} H. B. Zhao, K. J. Smith, Y. Fan, G. Lupke, A. Bhattacharya, S. D. Bader, M. Warusawithana, X. Zhai and J. N. Eckstein, \textit{Phys. Rev. Lett.} \textbf{100}, 117208 (2008)
\bibitem{metal insulator transition and its relation to magnetic structure} A. Bhattacharya, S. J. May, S. G. E. te Velthuis, M. Warusawithana, X. Zhai, Bin Jiang, J.-M. Zuo, M. R. Fitzsimmons, S. D. Bader and J. N. Eckstein, \textit{Phys. Rev. Lett.} \textbf{100}, 257203 (2008)








\bibitem{tricolor_growing} K. Lee, J. Lee and J. Kim, \textit{J. Korean Phys. Soc.} \textbf{46}, 112 (2005)

\bibitem{NdMnO3} J. Hemberger, M. Brando, R. Wehn, V. Yu. Ivanov, A. A. Mukhin, A. M. Balbashov and A. Loidl, \textit{Phys. Rev. B.} \textbf{69}, 064418 (2004)

\bibitem{Lattice effects on the magnetoresistance in doped LaMnO3} H. Y. Hwang, S-W. Cheong, P.G Radaelli, M. Marezio and B. Batlogg, \textit{Phys. Rev. Lett.} \textbf{75}, 914 (1995).

\bibitem{scaling} T. Sasagawa, P. K. Mang, O. P. Vajk, A. Kapitulnik and M. Greven,  \textit{Phys. Rev. B} \textbf{66}, 184512 (2002)
\bibitem{CdInCrS} E. Vincent and J. Hammann, \textit{J. Appl. C.} : Solid state Phys. \textbf{20}, 2659 (1987)

\bibitem{1-n} R. V. Chamberlin, \textit{J. Appl. Phys.} \textbf{57} 3377 (1985)
\bibitem{SG_aging}   D. X. Li, T. Yamamura, S. Nimori, K. Yubuta and Y. Shiokawa, \textit{Appl. Phy. Lett.} \textbf{87}, 142505 (2005)
\bibitem{Enhanced ordering temperatures in antiferromagnetic manganite superlattices} S. J. May, P. J. Ryan, J. L. Robertson, J.-W. Kim, T. S. Santos, E. Karapetrova, J. L. Zarestky, X. Zhai, S. G. E. te Velthuis, J. N. Eckstein, S. D. Bader and  A. Bhattacharya, \textit{Nature Materials} \textbf{8}, 892 (2009)
\bibitem{LFOLMO} K. Ueda, H. Tabata, and T. Kawai, \textit{Phys. Rev. B}, \textbf{60} R12561 (1999)
\bibitem{interface_SG_Fe_FeOxide}  L. Del Bianco, D. Fiorani, A. M. Testa, E. Bonetti, L. Savini and S. Signoretti  \textit{Phys. Rev. B}, \textbf{66}, 174418 (2002)



















\end{thebibliography}
\end{document}